\documentclass{ptephy_v1}

\usepackage{graphicx}
\usepackage{bm}
\graphicspath{{./}}
\usepackage{subfig}
\usepackage[separate-uncertainty=true,multi-part-units=single]{siunitx}

\begin{document}

\title{Optical shape analysis based on discrete Fourier transform and second order moment calculation of the brightness distribution for the detection of sub-micron range low energy tracks}


\author{Atsuhiro Umemoto}

\author[2]{Tatsuhiro Naka}

\author[3] {Toshiyuki Nakano}

\author[1]{Ryuta Kobayashi}

\author[2]{Takuya Shiraishi}

\author[4]{Takashi Asada}

\affil[1]{Graduate School of Science, Nagoya University, Furo-cho, Chikusa-ku, Nagoya,
464-8602, Japan\email{umemoto@flab.phys.nagoya-u.ac.jp}}
\affil[2]{Toho University, 2-2-1 Miyama, Funabashi, Chiba, Japan 274-8510}
\affil[3]{Kobayashi-Maskawa Institute, Nagoya University, Furo-cho, Nagoya, Aichi, Japan 464-8601}

\affil[4]{Laboratori Nazionali del Gran Sasso Via G. Acitelli, 22 67100 Assergi L'Aquila, Italy}


\begin{abstract}
~~ To recognize sub-micron range low energy tracks recorded in a super fine grained nuclear emulsion (Nano Imaging Tracker), an elliptical fitting method was devised to analyze anisotropic images taken by an optical microscope. In this paper, we will report on this newly developed method using discrete Fourier transform and second-order moment analysis of the brightness distribution. We succeeded in lowering the ellipticity threshold, thereby improving the detection efficiency and angular resolution. Notably, the success of detecting carbon 30 keV tracks is the first such achievement in the world, where the incident direction of carbon 30 keV ions was determined with an accuracy of 41$^\circ$ and an efficiency of 1.7 $\pm$ 0.1$\%$.
\end{abstract}
\subjectindex{H20, H21, C44}

\maketitle

\section{Introduction}

~~The spatial resolution of a particle dtracking detector determines the accuracy of elementary process measurements such as particle decay or directional measurement detected by its geometric construction. The super fine-grained emulsion Nano Imaging Tracker (NIT) is a solid-state tracking detector with the world's finest spatial resolution of several tens of nanometers, enabling sub-micron accurate particle tracking~\cite{NIT}.
Owing to its resolution, NIT will be able to realize directional dark-matter detection experiments, with the potential to verify the existence of Weakly Interacting Massive Particles (WIMPs) with low background leakage~\cite{NEWS}\cite{Antonia}.
The detection of the directional anisotropy of recoil nuclei will make it possible to determine the dark-matter velocity distribution in the Milky Way galaxy~\cite{nagao}.

To realize a read-out system with high spatial resolution compatible with the NIT resolution, we have been developing an automated read-out system based on an epi-illumination optical microscope~\cite{PTS2}, and have proposed a combination of analysis methods using super-resolution plasmonic imaging~\cite{plasmon} and elliptical analysis~\cite{elli1}\cite{PTS2}. 
In the elliptical analysis, the track direction and the track length are obtained from the anisotropic shape of the optical image.
We need to read out a large area of NIT, as expected in the case of directional dark-matter search experiments. Elliptical analysis is suitable because it requires no extra system except for a normal microscope.
In this paper, we will demonstrate a new elliptical analysis algorithm using a frequency space filter in the discrete Fourier transform (DFT) method. As described below, we succeeded in reducing the pixel digitization effect on the read-out image, and improved the detection efficiency and angular resolution of the sub-micron tracks. 

\section{Experimental setup}
\subsection{Super-fine grained nuclear emulsion}
\begin{figure}[htbp]
 \begin{center}
  \includegraphics[width=120mm]{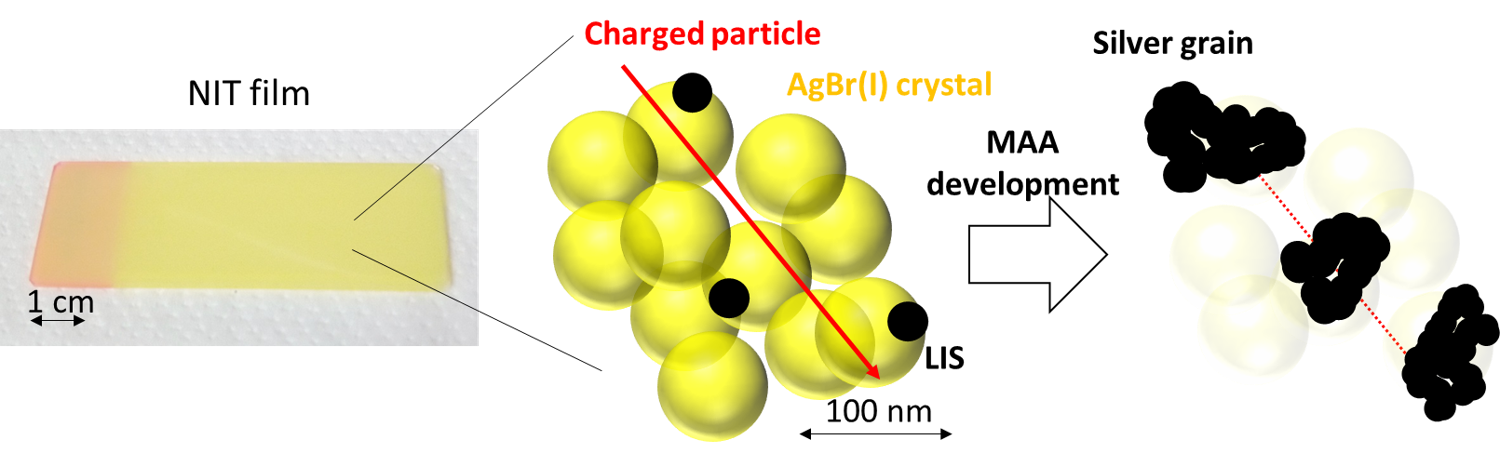}
  \caption{Photograph of an NIT film and a schematic view of the creation of track-grains from the MAA development. In this study, an NIT with a crystal size of 75.3 nm was used.}
 \label{fig:NITdetect}
 \end{center}
\end{figure}

~~NIT is a super fine-grained nuclear emulsion developed to detect sub-micron long charged particles. A schematic view of the particle detection in NIT is shown in Fig.~\ref{fig:NITdetect}.
The detector elements are silver bromide nano-crystals with a small amount of iodine additives (AgBr(I)) dispersed in a gelatin binder. The electrons in the AgBr(I) excited by ionization of charged particles create a silver core referred to as ”Latent Image Specs” (LIS) by binding with interstitial silver ions sequentially. Through the image development process, the LIS grow to silver grains with sizes that can be observed under an optical microscope. The track of a charged particle then emerges as a sequence of silver grains. Hereafter, LIS after development is referred to as a ”track-grain”. 

The AgBr(I) size and its number density are controllable. 
In this study, NIT with a crystal size of 75.3 $\pm$ 9.2 nm and an average crystal number density of 6.9 crystals/$\mu$m is used. NIT was coated on a side of slide glass to form a 1 $\mu$m thickness NIT layer. Because at least two grains are required to be recognized as a track, in principle more than 145 nm long tracks (inverse of crystal number density) are able to be detected as tracks.
Practically, the detection capability of NIT is determined by the position distribution of AgBr(I) crystals, ionization (i.e., dE/dx) of incoming particles, crystal sensitivity and development method~\cite{naka}.

In this study,  we used the ”metol-ascorbic-acid” (MAA) recipe of a standard chemical developer~\cite{PSA} which generates track-grains with complicated filament structures as shown in Fig.~\ref{fig:NITdetect}. The MAA was processed at 5$^\circ \mathrm{C}$ for 10 min and had enough sensitivity to produce the track-grains of low-velocity ions.
Although the complicated filament structure of the track-grains causes a non-uniform optical response like the brightness, it still has the advantage of suppressing production of silver grains, referred to as fog, which are randomly processed noises separately from tracks. 

\subsection{Sample for performance evaluation}
\subsubsection{Low-velocity carbon ion implantation}$\newline$
 ~~~~ To mimic the recoil ions for the evaluation of elliptical analysis quality, low-velocity carbon ions accelerated by an ion-implantation system for semiconductor device manufacturing are exposed to the NIT.
The setup enables irradiation with carbon ions with a monochromatic energy ($\pm$ 1 keV) and a small deviation in incident angle ($\leq$ 10 mrad). The carbon ion act as a calibration sample for carbon recoil in gelatin induced by WIMPs.
The minimum acceleration voltage of the equipment was 30 kV. Here, carbon ion energies of 30, 60, and 100 keV were selected and exposed to the NIT with a tilt angle of 10$^\circ$ parallel to the NIT surface as shown in Fig.~\ref{fig:C100}(a).

The kinetic energies 30, 60, and 100 keV correspond to the maximum recoil energy of carbon in the case of WIMPs masses of 10, 25, and 50 GeV/$c^2$, respectively. The maximum track lengths of carbon ions with energies of 30, 60, and 100 keV are estimated to be 220, 360, and 550 nm, respectively~\cite{SRIM}. As these lengths are greater than the two-crystal distance as described, these ions should produce tracks by reacting with several crystals.

\subsubsection{Silver nano particles}$\newline$
 ~~~~ Samples of spherical Ag particles (product number 730807 manufactured by SIGMA-ALDRICH Co., Ltd., diameter 40 $\pm$ 4 nm ) were prepared by dispersing in a thin gelatin layer.
In sub-micron track detection, each spherical particle must be recognized as a sphere. The optical image of an Ag particle is regarded to be the image of a point source, because the particle size is at least ten times smaller than the wavelength used.

This sample contains not only spherical Ag particles but also impurities in gelatin and outside contamination. As the spherical particles make a chance coincidence (CC) by approaching each other, we prepared two samples with different particle number densities to estimate the effects of CC and contamination on the analysis.

\subsection{Read-out system}
\begin{figure}[htbp]
 \begin{center}
  \includegraphics[width=60mm]{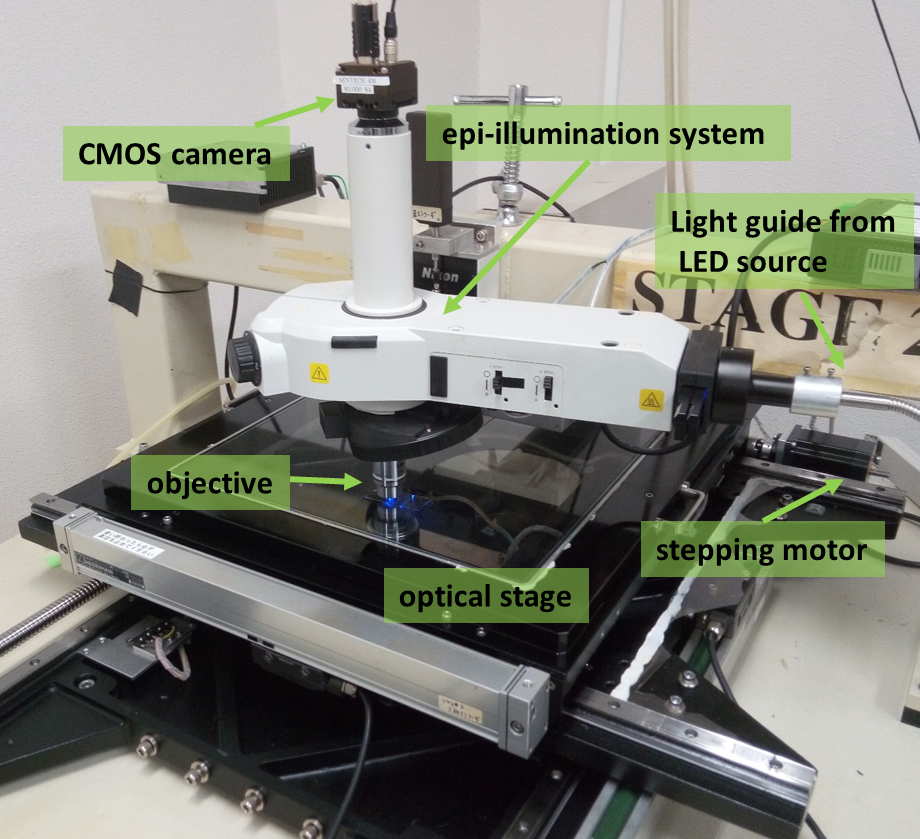}
  \caption{Photograph of the PTS2 setup on the basis of an epi-illumination optical microscope.}
 \label{fig:PTS2}
 \end{center}
\end{figure}

~~We have developed an automated read-out system named PTS2 based on an epi-illumination optical microscope (Fig.~\ref{fig:PTS2})\cite{PTS2}. The optical system is equipped with an oil immersion objective lens ($\times$ 100, numerical aperture (NA) 1.45), and a CMOS camera (2048 $\times$ 2048 pixels, 160 frames/s read-out speed), an LED light source with wavelength 455 nm, and stepping motors controlling an $XYZ$ stage.
The stage position is controlled by a motion board on a PC. The image from the camera is captured by an image grabber on the PC with object digitization of 55 nm/pixel. 
Typically, 96 images of 30 $\mu$m thickness are taken by changing the $Z$-coordinate, and the in-focus layer is identified as the brightest image.

\begin{figure}[htbp]
 \begin{center}
  \includegraphics[width=135mm]{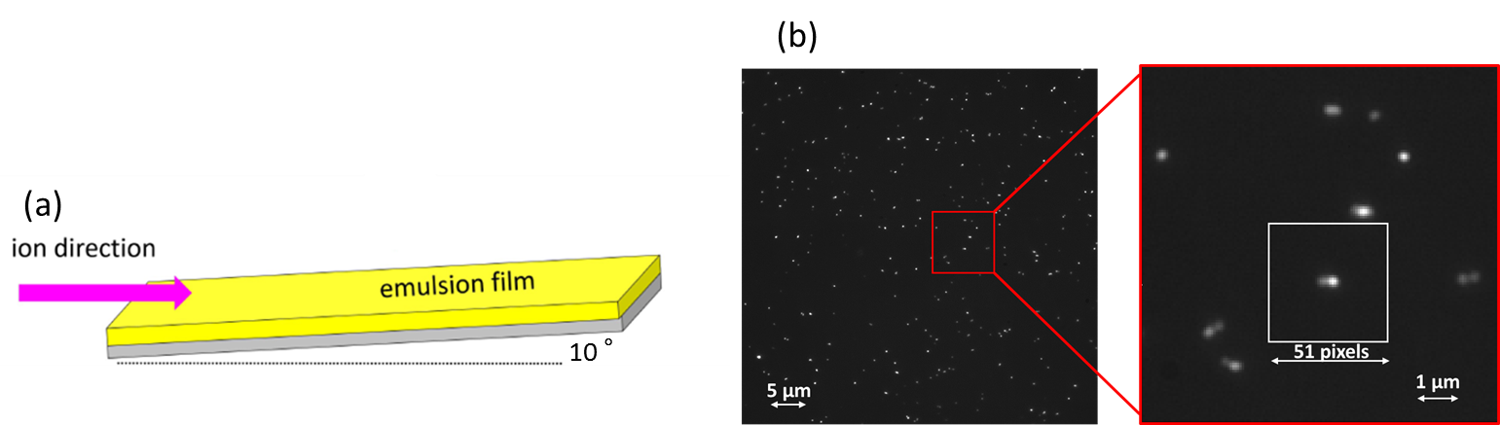}
  \caption{(a) : NIT installation during ion implantation. (b): Optical image of carbon 100 keV tracks recorded in NIT.}
 \label{fig:C100}
 \end{center}
\end{figure}

Fig.~\ref{fig:C100}(b) shows an optical image of carbon 100 keV track taken by PTS2. The horizontal direction of Fig.~\ref{fig:C100}(b) is the beam direction, and each event is a track of carbon ions. The point spread function (PSF) of a microscope image is determined by the wavelength and NA of the objective lens, and is measured to be 214 $\pm$ 8 nm in the case of PTS2.
This resolution is not sufficient to resolve each track-grain in the carbon 100 keV track, but the optical image shows an anisotropic shape slightly extended in the beam direction. Therefore, shape recognition of the optical image can extract information on the sub-micron track. 

We demonstrated the performance of the elliptical analysis technique known as contour fitting in a previous paper (hereafter this is referred to as the old method)~\cite{PTS2}.
As shown in the next section, this method has a problem caused by pixel digitization, because it performs the image processing in pixel space. In particular, the calculated elliptical parameters show a non-isotropic angular response with respect to the direction of the pixel coordinate. In addition, as the track information used to calculate the ellipticity is limited in a cluster defined by the contour extraction of the binarized image, a misrecognition of optical shape can be caused by the uncertainty in binarization.

To improve this situation, a new elliptical analysis algorithm using discrete Fourier transform (DFT) and second-order brightness moments for ellipticity calculations has been developed (hereafter referred to as the new method).
In the next chapter, we will report on the algorithm and compare the performance of the new method with the old method. 

\section{Elliptical shape analysis}
\subsection{Algorithm for elliptical shape detection}
~~The process scheme and the feature of each method will be described using a real optical image of a spherical silver particle taken by PTS2.
The processing results of each step are shown in Fig.~\ref{fig:imgprocess}. Fig.~\ref{fig:imgprocess}(a) is an original image of figure size is 51 $\times$ 51 pixels, and the brightness of each pixel is shown in color with resolution of 8 bits (256 steps).

\subsubsection{Contour fitting method (old method)}$\newline$
 ~~~ In the old method, elliptical parameters are calculated through contour extraction after recognizing the optical image as a cluster. The image processing filter is created in consideration of the luminous image size concentrated within about 6 $\times$ 6 pixels and its impulse response.
At first, a blur filter with a box size of 5 $\times$ 5 is applied for smoothing (Fig.~\ref{fig:imgprocess}(b)).
Then high-frequency components are then extracted by subtracting images before and after a low pass filter. The low pass filter, with cutoff frequency of 0.08 cycles/pixel and Hamming window function of 1 $\times$ 15 and 15 $\times$ 1, is applied in both X and Y directions.
Subsequently, the binarization process is applied to pixels with brightness greater than 1. The result is shown in Fig.~\ref{fig:imgprocess}(c). 
The expansion after shrinking the image (morphological opening process) is then applied to discriminate the remained spurious noise seen in Fig.~\ref{fig:imgprocess}(c). 

The shape of the obtained binarized cluster shown in Fig.~\ref{fig:imgprocess}(d) appears elliptical, although the original image is spherical.
The ellipticity obtained by the subsequent calculation is 1.20. This distortion has a dependence on the image divergence on the sub-pixels. 

\subsubsection{DFT and moment (new method)}
\begin{figure}[htbp]
 \begin{center}
   \includegraphics[width=150mm]{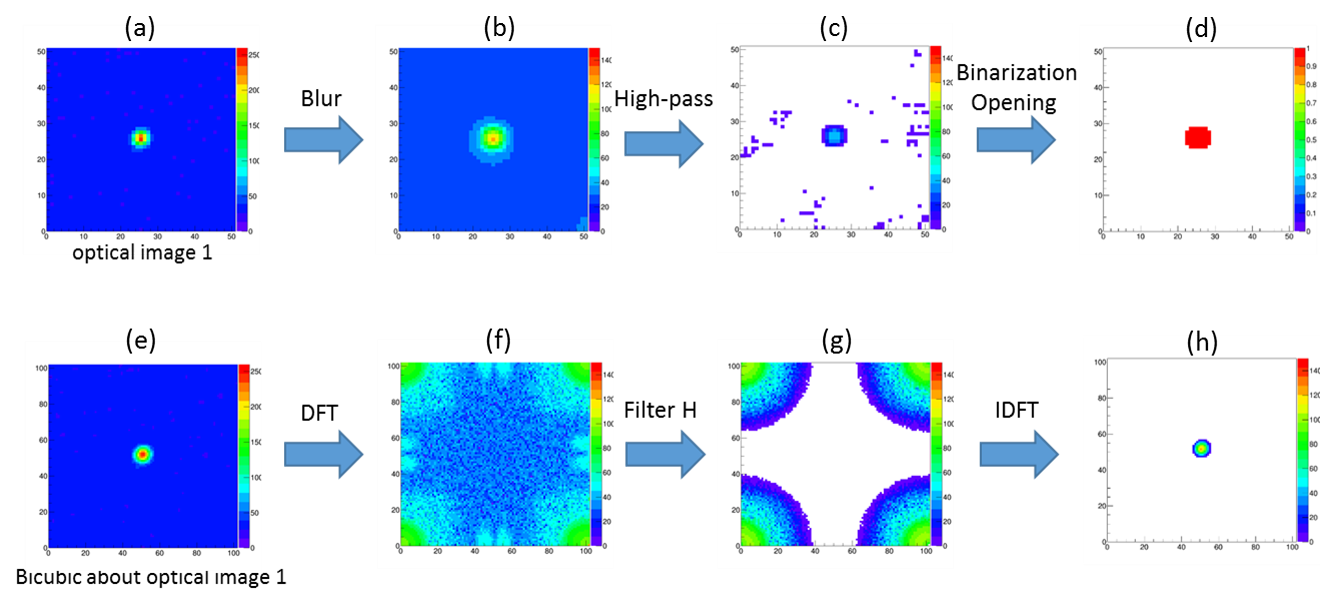}
 \caption{Image processing results of the elliptical analysis methods. The upper and lower figures correspond to the old and new methods, respectively.} 
 \label{fig:imgprocess}
 \end{center}
\end{figure}

$\newline$
 ~~~~ In the new method, the original image is expanded to 102 $\times$ 102 pixels by bicubic interpolation as shown in Fig.~\ref{fig:imgprocess}(e), to extend the wave number space and reduce the anisotropic effects in the following processes. 
The expanded image is converted to frequency spectrum space by a discrete Fourier transform (DFT).

\begin{figure}[htbp]
 \begin{center}
   \includegraphics[width=70mm]{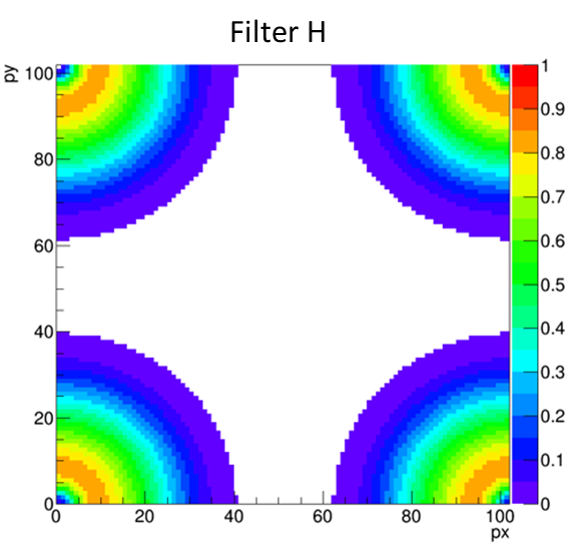}
 \caption{Frequency filter used in the step of Fig.~\ref{fig:imgprocess} (f) to (g). The low-frequency components of the event are  efficiently transmitted, so that isotropic processing can be performed in the wave number space. The color indicates the transmittance.} 
 \label{fig:FilterH}
 \end{center}
\end{figure}

In Fig.~\ref{fig:imgprocess}(f), lower-frequency components appear in the four corners. Because the spherical image in Fig.~\ref{fig:imgprocess}(a) is spread to 6 $\times$ 6 pixels as mentioned (i.e. not in the high frequency region), a frequency filter shown in Fig.~\ref{fig:FilterH} is developed and applied to Fig.~\ref{fig:imgprocess}(f). The filtered spectrum shown in Fig.~\ref{fig:imgprocess}(g) is inversely converted to the space figure as shown in Fig.~\ref{fig:imgprocess}(h). In this conversion, only components for which brightness is greater than 3 are used to reduce the noise. 
In the optical image after processing (Fig.~\ref{fig:imgprocess}(h)), it can be seen that the luminous component of the original image is extracted well, and the shape looks spherical.

In the new method, we also developed a method to use the brightness information, i.e., the same process as the mechanical inertia moment calculation in two dimensions. 
Using the distance of each pixel from the barycenter $(x_{ic}, y_{ic})$ and brightness of each pixel $(B_{ic})$, an inertia matrix can be defined as shown in equation (\ref{eq:moment}). By solving the intrinsic equation, intrinsic values $(a_{11}, a_{22}, a_{33})$ and intrinsic vectors $(\bm{e}_1, \bm{e}_2, \bm{e}_3)$ are obtained.
The direction of the first main axis is the major axis of the elliptical shape. By using the ratio of $a_{11}$ and $a_{22}$ to the main inertia moment $a_{33}$, we can derive the minor as 2$\times$$(a_{11}/a_{33})^{\frac{1}{2}}$ and the major as 2$\times$$(a_{22}/a_{33})^{\frac{1}{2}}$ of the ellipse. These parameters correspond to a full width of brightness distribution in each axis of the minor and major. The ellipticity, defined as the major divided by the minor, is then obtained.
The ellipticity of Fig.~\ref{fig:imgprocess}(h) is calculated to be 1.06, therefore, a better result is obtained by the new method compared to the old method. 

\newpage

\begin{equation}
I=\sum_{i=0}^{n p i x e l} B_{i c}\left[\begin{array}{ccc}
{y_{i c}^{2}} & {-y_{i c} x_{i c}} & {0} \\
{-x_{i c} y_{i c}} & {x_{i c}^{2}} & {0} \\
{0} & {0} & {x_{i c}^{2}+y_{i c}^{2}}
\end{array}\right]=\left(\begin{array}{ccc}
{a_{11}} & {0} & {0} \\
{0} & {a_{22}} & {0} \\
{0} & {0} & {a_{33}}
\end{array}\right)\left(\begin{array}{c}
{e_{1}} \\
{e_{2}} \\
{e_{3}}
\end{array}\right)
 \label{eq:moment}
\end{equation}
\clearpage

\subsection{Performance for spherical nanoparticle reduction}
\begin{figure}[htbp]
 \begin{center}
   \includegraphics[width=120mm]{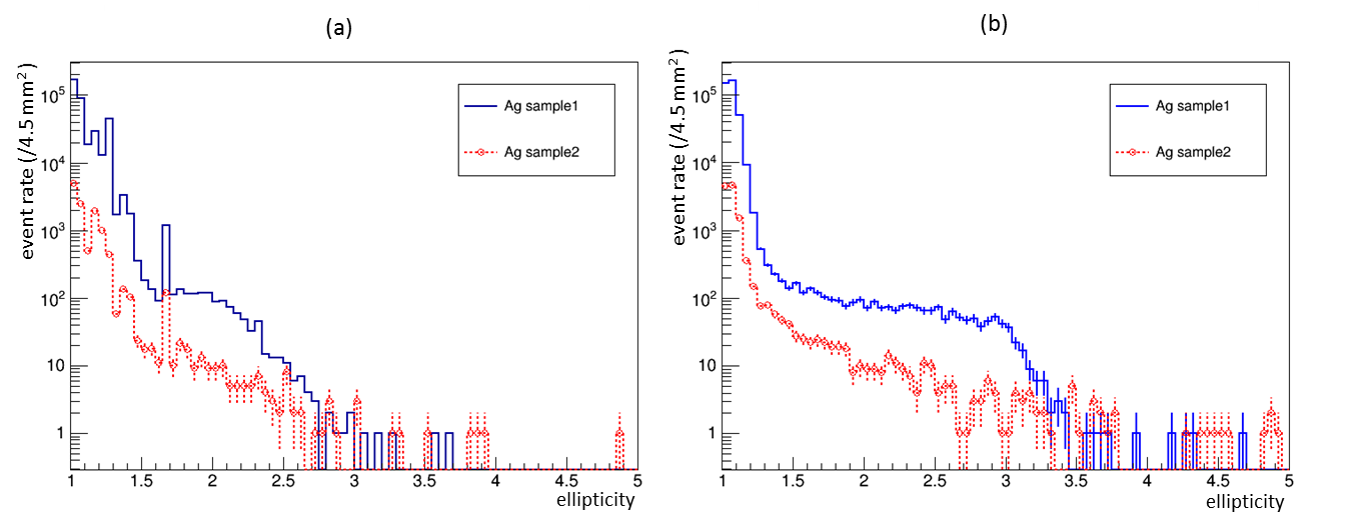}
 \caption{Elipticity distribution of the Ag sample. (a) by the old method, (b) by the new method. The blue plot with a solid line is the case of sample 1(high density) and the red plot with a dotted line is sample 2(low density).} 
 \label{fig:Agelli}
 \end{center}
\end{figure}

~~To evaluate the recognition capability of spherical particles by each method statistically, two samples were used with different particle densities of (0.82 $\pm$ 0.10) $\times$ $10^7$/$cm^2$ (Ag sample1), close to the densities of the ion implantation samples used in the next section 3-3, and of (0.03 $\pm$ 0.02) $\times$ $10^7$/$cm^2$ (Ag sample2), to estimate the CC rate and contamination.
Figure.~\ref{fig:Agelli} shows the ellipticity distribution of Ag samples 1 and 2 with 4.5 mm$^2$ scanning obtained by the old method (Fig.~\ref{fig:Agelli}(a)) and by the new method (Fig.~\ref{fig:Agelli}(b)), respectively. 

In the old method, spikes at certain ellipticity values appear in the distribution even though Fig.~\ref{fig:Agelli} is expressed on a logarithmic scale.
In contrast, the distribution width around ellipticity 1 becomes thinner in the new method, and the event rate decreases rapidly with an increase in ellipticity. In addition, it can be seen that the ellipticity distribution is flat between 1.5 and 3 in Ag sample 1, although there is a decreasing tendency in sample 2. This difference can be explained by the existence of the CC of two particles approaching within a distance less than 300 nm. 
From the distribution shown in Fig.~\ref{fig:Agelli}(b), the CC appears to start at ellipticity 1.3. The estimated CC rate of Ag sample1 is $\sim$ 4000 from Poisson calculation, which is not far from the detected number of 3290 with an ellipticity threshold of 1.3.
In the case of Ag sample 2, the CC rate is estimated to be 15, but 648 events are detected with an ellipticity threshold of 1.3. It can be said that the events with high ellipticity are dominated by CC in the case of Ag sample 1, and by contaminations in the case of sample 2.

\begin{figure}[htbp]
 \begin{center}
   \includegraphics[width=120mm]{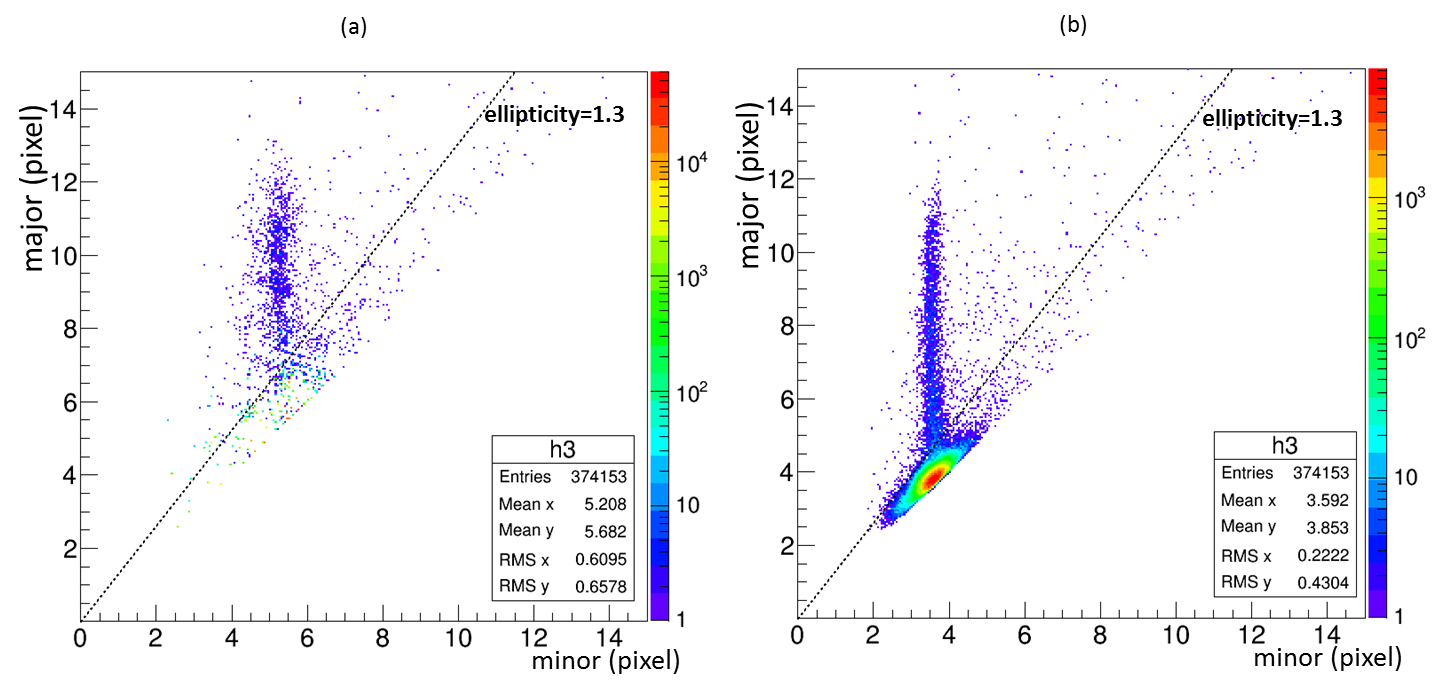}
 \caption{Major and minor distribution about Ag sample 1. (a) by the old method, (b) by the new method. The dashed lines in the figures show the case of ellipticity 1.3. } 
 \label{fig:Agminmaj}
 \end{center}
\end{figure}

Figure.~\ref{fig:Agminmaj} shows the distribution of the major and minor for elliptical recognition of Ag sample 1 by the old method (a) and by the new method (b).
The existence of the CC can be recognized as the vertical-line shape distributions at minor 5.5 in the case of the old method, and at minor 3.5 in the new method. To evaluate the sharpness of the distribution for a single spherical Ag particle, the CC events can be rejected by requesting that the ellipticity is smaller than 1.3. The events below this value are used for the evaluation of the spherical particles. In this region, the minor and major are respectively described as Gaussians of 3.59 $\pm$ 0.24 and 3.82 $\pm$ 0.22 by the new method, although they are digitized by the old method with an average minor and major of 5.25 $\pm$ 0.60 and, 5.67 $\pm$ 0.57, respectively.  

\begin{figure}[htbp]
 \begin{center}
   \includegraphics[width=120mm]{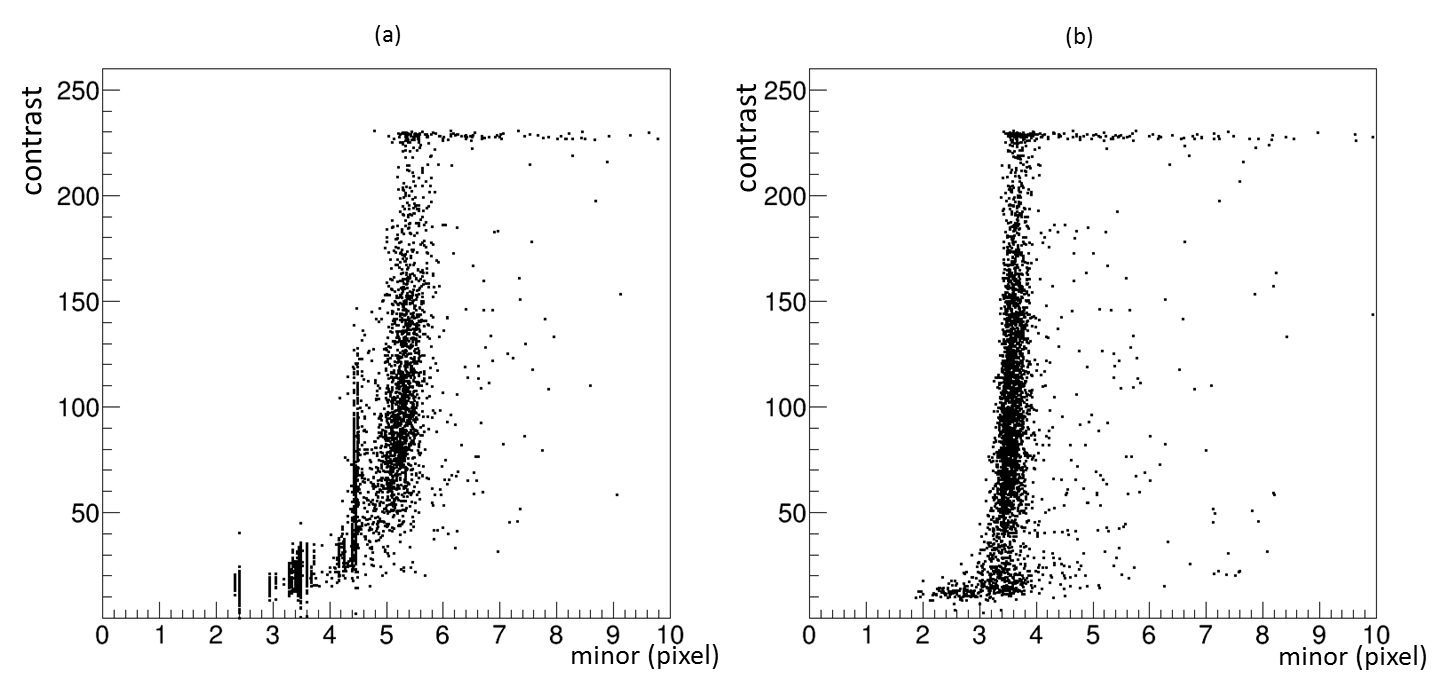}
 \caption{Contrast correlation with the minor in Ag sample 1. (a) by the old method, (b) by the new method. Selection of events with an ellipticity of 1.3 or more is applied in both methods.} 
 \label{fig:Agcontrast}
 \end{center}
\end{figure}

Here we focus on the relation between the minor and the contrast (contrast is defined as the difference between maximum brightness of the event and the background (average 25)) to understand the high ellipticity events.
The contrast might affect ellipticity systematically. The lower it is, the worse recognition of the optical shape becomes.
Figure.~\ref{fig:Agcontrast} shows the relationship between the minor and the contrast of an event with an ellipticity greater than 1.3. The events, for which the contrast is approximately 230, contain saturated pixels. 
In the old method, there are vertical structures around minors of 2.5, 3.5 and 4.5 caused by the remaining pixel structure (i.e., image digitization). As those are the factors degrading the accuracy of the ellipticity analysis, the condition of minor $\geq$ 4.8 has been applied in the old method. Contrast is also a factor affecting the ellipticity performance even in the new method. In Fig.~\ref{fig:Agcontrast}(b), there appear to be two components, one is the events distributed vertically around minor 3 to 4, and the others are the events distributed horizontally starting from a minor of  2 and a contrast of 10, and the saturated events.
We applied the criterion of a contrast as 20 or more to cut the latter low-contrast events adding to the cut on the saturated events. The loss of events due to this contrast cut is 4.6$\%$ in Ag sample 1. 

The sharpness of the ellipticity distribution in the case of the spherical particles in As sample 1 was evaluated.
As events from CC and contamination are included, not all the events are spherical particles. Therefore, we set the threshold of ellipticity for the recognition of spherical particles to a minimum value, which suppresses the number of events to less than 1$\%$. 
In the analysis, 374,153 events are detected. Without the contrast cut, the thresholds are 1.41 and 1.26 in the case of the old and the new method, respectively. With the contrast cut, 356,765 events remain and the thresholds become 1.40 and 1.22, respectively. To follow the defined normal sequence of the old method used in the ion implantation case, as shown later, a minor cut is also applied, after which a total of 271,434 events remain and the threshold lowers to 1.27, which is still higher than in the case of the new method even with the additional loss of 27.5 $\%$. In conclusion, the new method shows much better spherical particle identification capability than the old method.

\clearpage

\subsection{Performance for low-velocity ions}
\subsubsection{Tracking performance for carbon 100 keV}$\newline$
  Using the carbon 100 keV sample, the track detection capability was evaluated for each elliptical analysis method.
The detected event number density of carbon 100 keV is (1.14 $\pm$ 0.06 )$\times$$10^7$/$cm^2$, and the reference fog sample has an event number density of (0.04 $\pm$ 0.04)$\times$$10^7$/$cm^2$. Therefore, most events detected in carbon 100 keV are ion signals.
Elliptical analysis was performed by each methods using 1.5 mm $\times$ 1.5 mm scanning data.

\begin{figure}[htbp]
 \begin{center}
   \includegraphics[width=85mm]{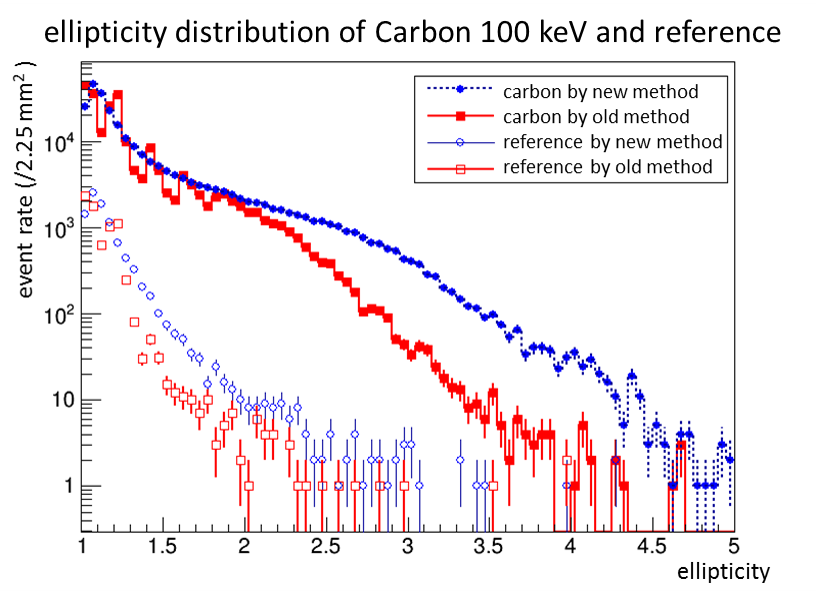}
 \caption{Ellipticity distribution of the carbon 100keV and the reference. The carbon and reference are plotted with dotted points and with open points, respectively. The blue histograms show the case treated by the new method, and the red by the old method.} 
 \label{fig:C100elli}
 \end{center}
\end{figure}

Figure.~\ref{fig:C100elli} shows the ellipticity distribution of the carbon 100 keV sample and the reference sample.
About carbon 100 keV, there are many events with high ellipticity and the dynamic range of the ellipticity appears to improve in the new method. Several spikes are seen in the distribution of the old method, such as the case of spherical Ag particles shown in Fig.~\ref{fig:Agelli}. 

\begin{figure}[htbp]
 \begin{center}
   \includegraphics[width=110mm]{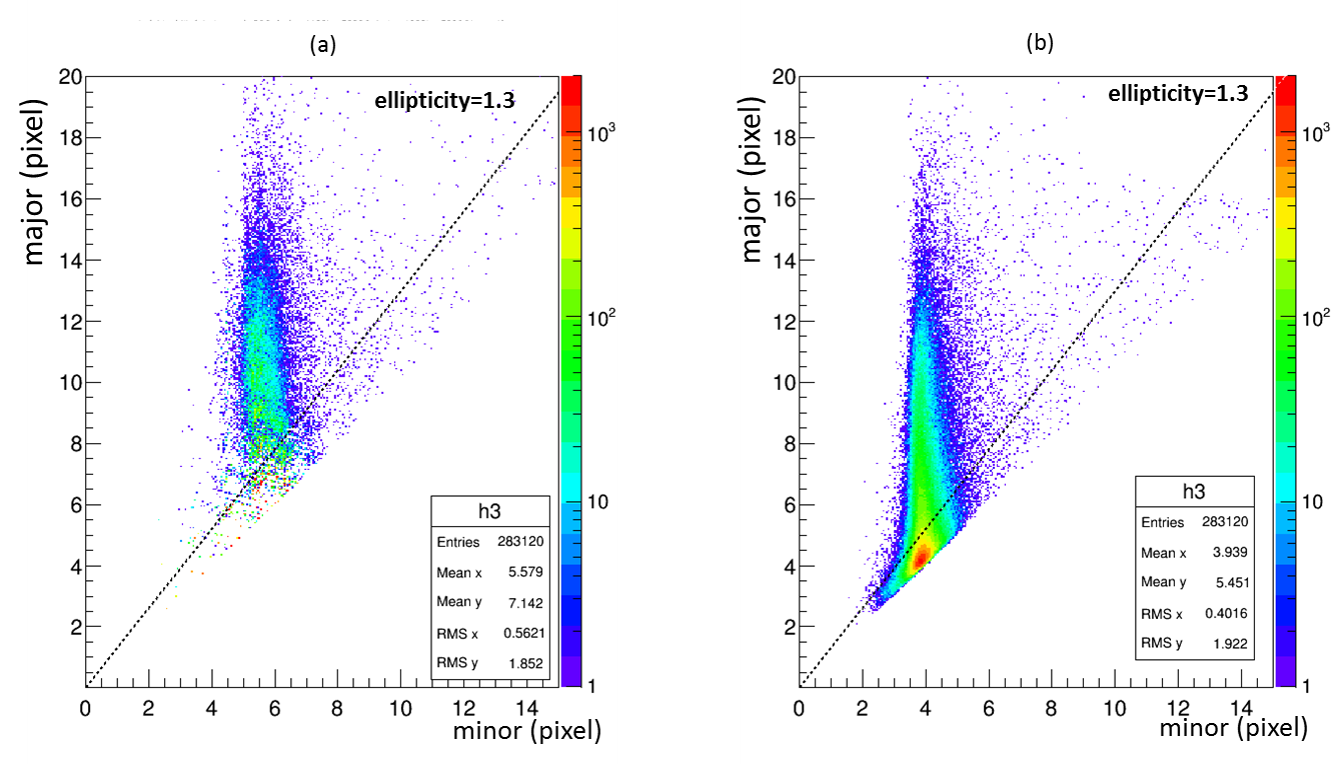}
 \caption{Major and minor distributions of Fig.~\ref{fig:C100elli} in the case of the carbon 100 keV sample. (a) by the old method, (b) by the new method.} 
 \label{fig:C100minmaj}
 \end{center}
\end{figure}

Figure.~\ref{fig:C100minmaj} shows the plot of the minor and major. The carbon 100 keV signals are vertically distributed at a minor of around 5.6 in the old method, and 3.9 in the new method. The center minor value of the vertical distribution appears to be slightly shifted to larger than in Ag sample 1. This may be caused by the difference in the silver grain size and structure developed from AgBr(I) with diameter 75.3 nm, which is larger than the Ag particle size of 40 nm in the Ag sample.
The digitizing events are still seen at an ellipticity of around 1.3 in the old methods. 
 
\begin{figure}[htbp]
 \begin{center}
   \includegraphics[width=110mm]{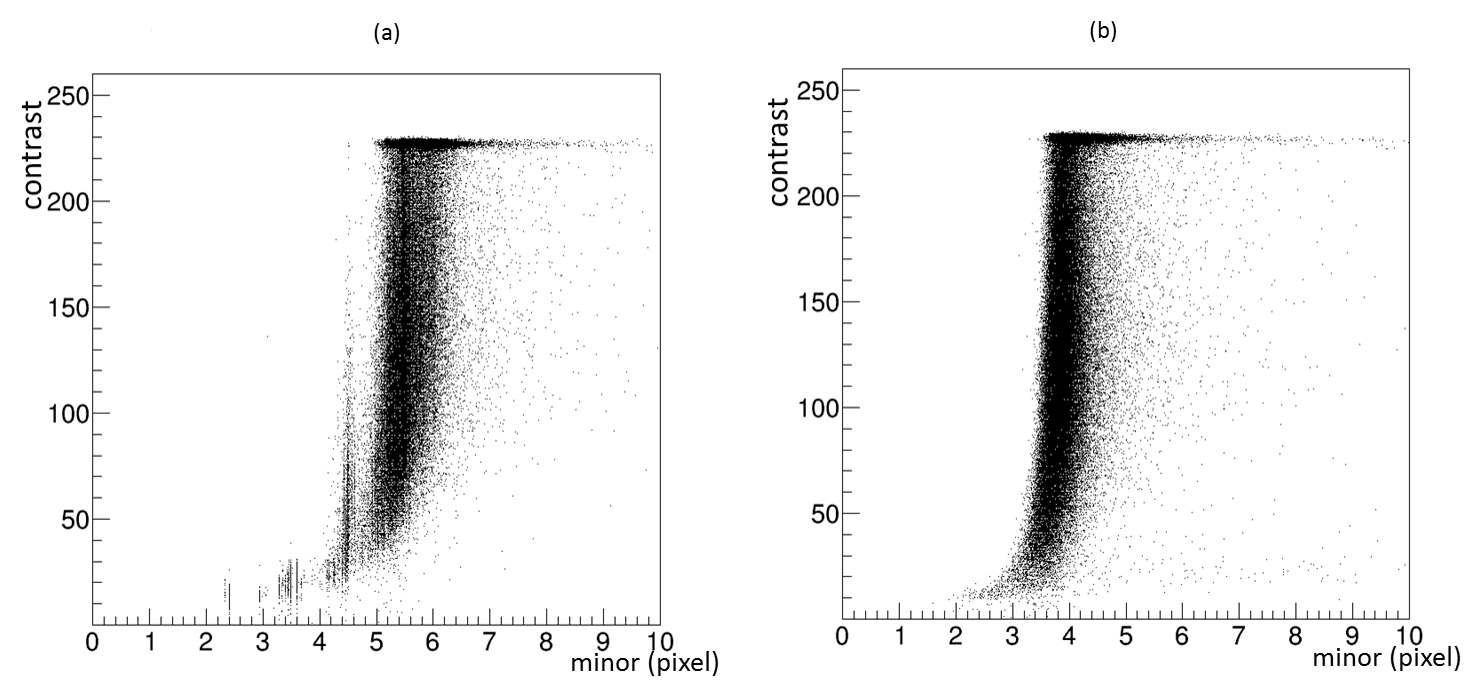}
 \caption{Contrast correlation with the minor in carbon 100 keV. Selection of events with an ellipticity of 1.3 or more is applied in both methods and (a) by the old method, (b) the new method.} 
 \label{fig:C100cont}
 \end{center}
\end{figure}

Figure.~\ref{fig:C100cont} shows the minor and contrast distributions about carbon 100 keV with ellipticity $\geq$ 1.3 selected.
As in the case of Ag sample analysis, a cut on minors less than 4.8 was applied in the old method. In addition, a cut on contrast less than 20 and a cut of the saturated events were applied in both methods to evaluate the signal. 
A total of 283,120 events were detected, of which 238,247 (84$\%$) remained after the contrast cut. Of the cut events, 40,650 (14$\%$) were by saturation and 4223 (1.5$\%$) by low-contrast cut. As the saturated events are the real tracks, improvement to the contrast treatment in the development process is needed in the future to improve the detection efficiency. For the utilization of the old method, the effective events number decreases to 219,332 (77$\%$) after the extra minor cut. 

To evaluate the uniformity of the direction determination, the sample was rotated manually on the PTS2 stage so that the beam direction was set to be 90$^\circ$, 135$^\circ$, 180$^\circ$ (0$^\circ$) on the angle space of the elliptical analysis.
The events after the brightness cut are used in the new method, and the events with a further minor cut are used in the old method. 
As the scanning zone was fixed to be 0.6 mm in $X$ and 0.6 mm in $Y$ on the read-out stage, the scanning area on the film was not the same in each rotated case. However, the number of detected events is consistent within a range of 10$\%$. The horizontal direction of the image is defined as 90$^\circ$, and the vertical direction is defined as 0$^\circ$ or 180$^\circ$.

\begin{figure}[!htbp]
 \begin{minipage}[b]{0.96 \linewidth}
  \centering
  \includegraphics[width=120mm]{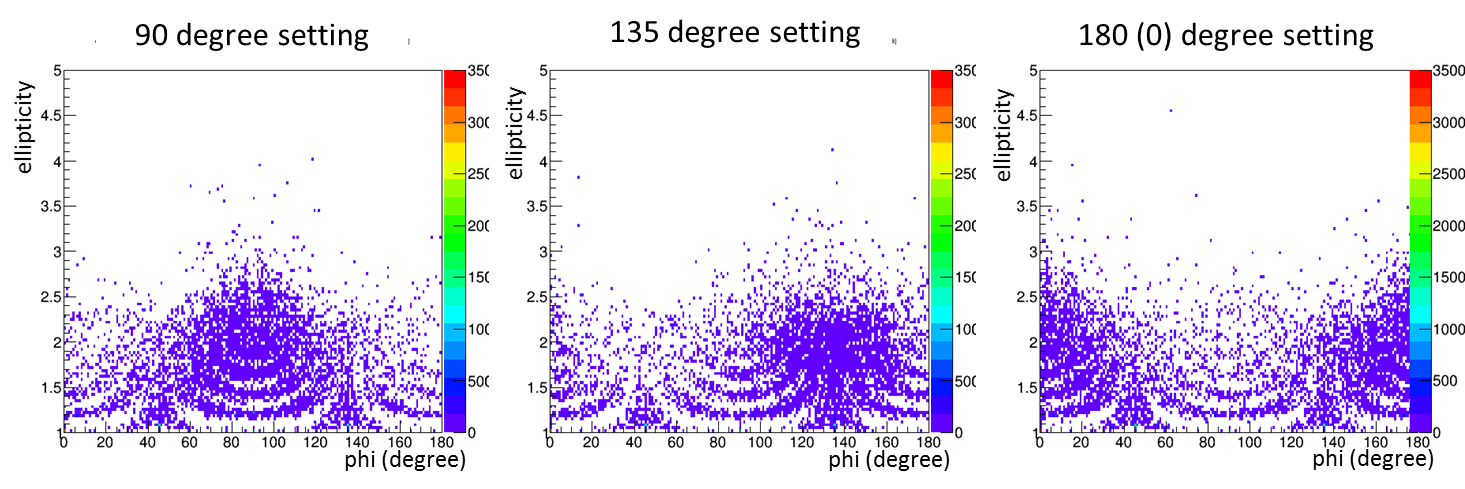}
  \end{minipage}\\
 \begin{minipage}[b]{0.96\linewidth}
  \centering
  \includegraphics[width=120mm]{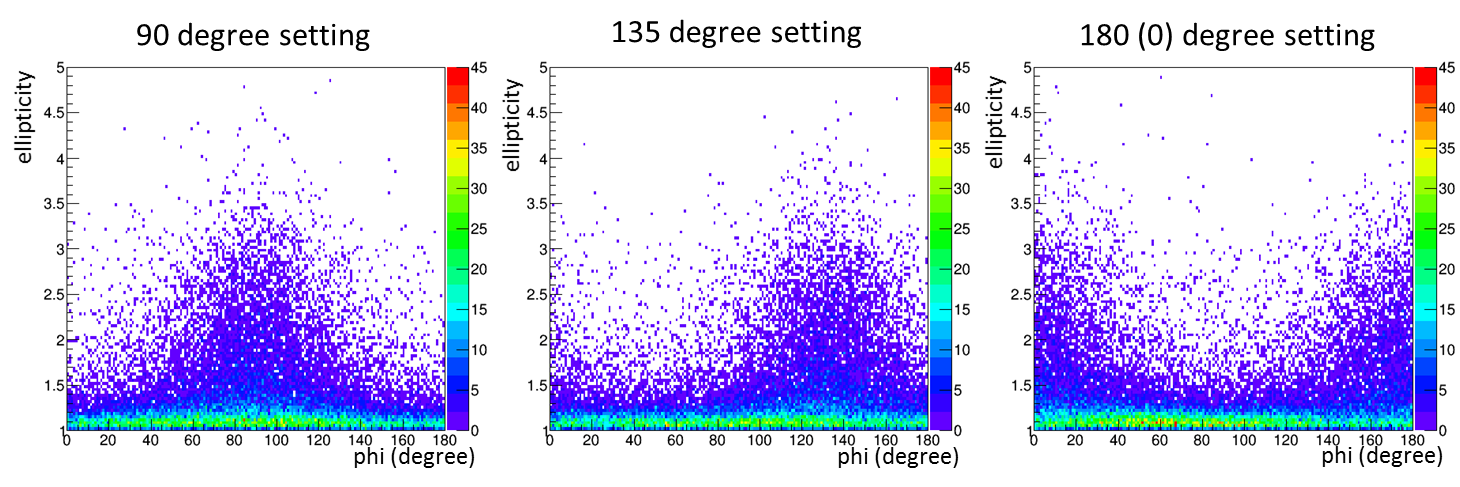}
 \end{minipage}
\caption{ Plots of the reconstructed ellipticity and angle in the case of carbon 100 keV. From left to right, the case of 90$^\circ$, 135$^\circ$, and 180$^\circ$ of the sample rotation. The upper figures show the case of the old method, and the lower are the new method.}\label{fig:C100_angle_elli}
\end{figure}

Fig.~\ref{fig:C100_angle_elli} shows the plots of the angle and the ellipticity of the reconstructed events in each method. In the old method, fixed striped patterns are seen in the distribution, i.e., the pixel digitization effects remain.
The position of the event concentration is shifted by the sample rotation as expected, but because of the existence of the fixed pattern, the distribution of 135$^\circ$ does not overlap with the distribution of 90$^\circ$ by a 45$^\circ$ rotation.
In contrast, there is no such fixed pattern in the results obtained by the new method. 

\begin{figure}[htbp]
 \begin{center}
   \includegraphics[width=120mm]{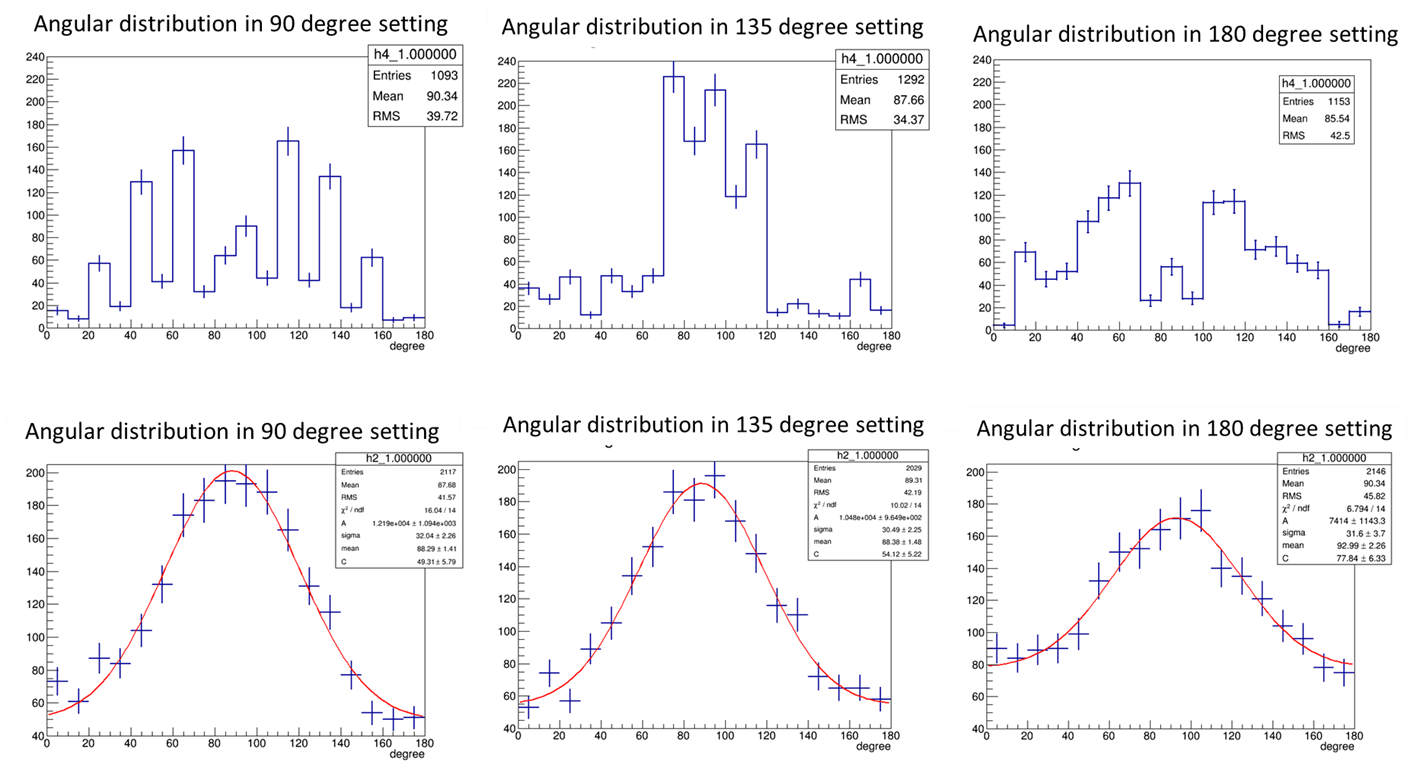}
 \caption{Angle distribution of the events for which reconstructed ellipticity is between 1.3 and 1.4. Upper and lower figures are the cases of the old method and the new method, respectively. In the figure expression, the rotation of the sample is corrected so that the beam direction is at 90$^\circ$.} 
 \label{fig:C100_e1314}
 \end{center}
\end{figure}

To evaluate the dependence of the angular resolution on the ellipticity, the plot shown in Fig.~\ref{fig:C100_angle_elli} is sub-divided with ellipticity intervals of 0.1, i.e., ellipticity intervals 1.2-1.3, 1.3-1.4, etc., up to 2.9-3.0. In Fig.~\ref{fig:C100_e1314}, the case of the ellipticity interval 1.3-1.4 is shown. The digitization effect is clear in the old method compared to the smoother result obtained by the new method. In addition, the shape of the distribution for a different setting angle is clearly different in the case of the old method, but not as noticeably different in the case of the new method. The difference between setting angle on the new method come from the micro-vibration of stage and the aberration of the optics mentioned in next section.

\begin{figure}[htbp]
 \begin{center}
   \includegraphics[width=120mm]{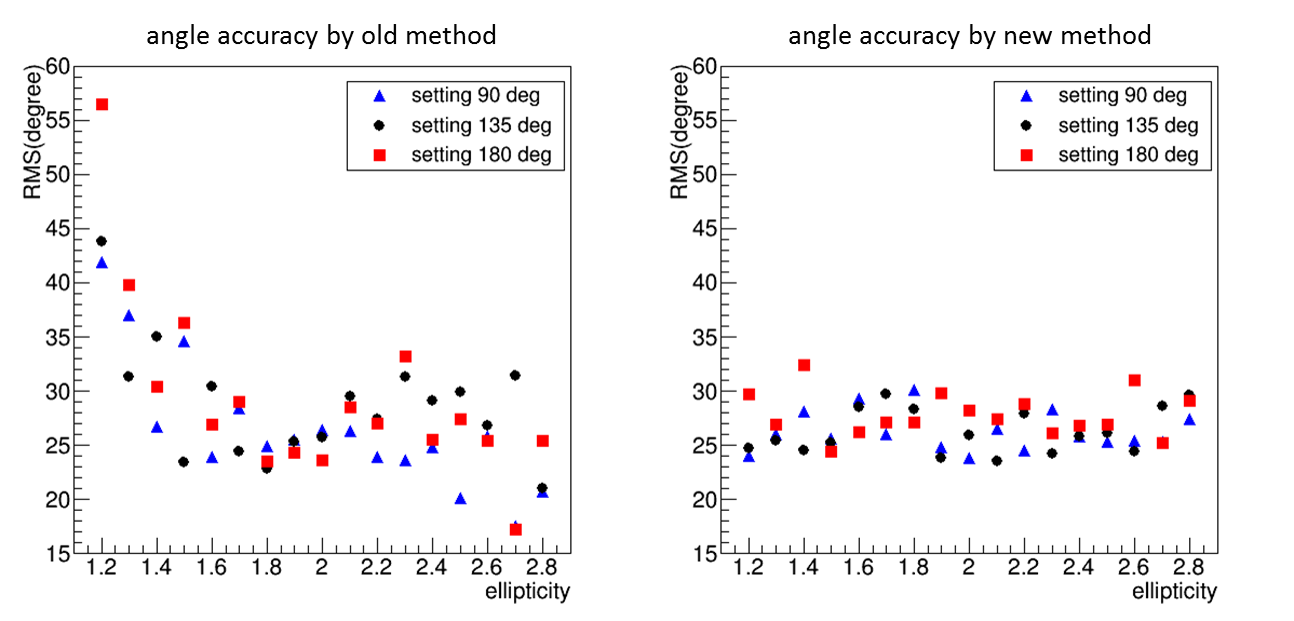}
 \caption{Dependence of angular resolution on ellipticity. Left and right figures are the cases of the old method and the new method, respectively. Note that the expected RMS of the flat distribution is (180/120)$^{0.5}$ = 52$^\circ$ } 
 \label{fig:C100_RMS}
 \end{center}
\end{figure}

The angular resolution, which is defined as the root mean square (RMS) of the deviation of the reconstructed angle from the real rotation angle, is shown in Fig.~\ref{fig:C100_RMS}. 
In the old method, the RMS depends on the ellipticity and shows large differences between setting angles. In comparison, the new method shows almost no dependence on either the ellipticity or the setting angle. 
In the old method, we require an ellipticity threshold of more than 1.6 to obtain the same deviation of RMS as that seen in the new method with ellipticity of more than 1.2.

The average RMS of the old method over the ellipticity of the discussed region is 27 $\pm$ 5, 29 $\pm$ 3, and 31 $\pm$ 9 degree for the setting position of 90$^\circ$, 135$^\circ$, and 180$^\circ$, respectively. In the case of the new method, it is 27 $\pm$ 2, 26 $\pm$ 2, and 28 $\pm$ 2 degree, respectively. The value of 27$^\circ$ is not far from the spread caused by the multiple Coulomb scatterings of incident ions in NIT. The improvement by the new method is clear in the low-ellipticity region.

\newpage

We can set the track detection threshold on the ellipticity when it is judged that the beam incident angle is reconstructed well at that threshold ellipticity. The detection efficiency ($R$) is defined as the ratio of the number of detected events above the threshold ($N_{det}$) to the total number of events ($N_{all}$). Similarly, the detected events about the reference are denoted by ($N_{det ref}$) and ($N_{all ref}$).
As there is a background in the reference sample even without beam exposure, a small correction is needed of 
$R$ = ($N_{det} -N_{det ref}) / (N_{all}-N_{all ref})$. The results in the case of ellipticity threshold 1.3 are shown in Table \ref{tb:C100_eff}. The detection efficiency of carbon 100 keV is about 30$\%$ via the new method, which is approximately 1.5 times higher than that obtained by the old method. 

\begin{table}[htbp]
\centering
\begin{tabular}{|c|c|c|c|c|c|c|}
\hline
analysis method & setting angle & $N_{all}$ & $N_{det}$ &$N_{all ref}$ &  $N_{det ref}$ & $R$\\ 
\hline
& 90$^\circ$& 36621 & 6853&1393 &  40 & 19.3($\%$)\\
old method& 135$^\circ$ & 36310 & 6698 &1393 &  40 & 18.9($\%$)\\
& 180$^\circ$ & 38678 & 6623 &1393 &  40 & 18.7($\%$)\\
\hline
& 90$^\circ$ & 36621 & 10810 & 1393 & 56 & 30.5($\%$)\\
new method & 135$^\circ$ & 36310 & 10441 & 1393 & 56 & 29.7($\%$)\\
& 180$^\circ$ & 38678 & 10803 & 1393 & 56 & 28.8($\%$)\\
\hline
\end{tabular}
\caption{Detection efficiency of the 100 keV samples by the old and the new method. The ellipticity threshold was 1.3.}
\label{tb:C100_eff}
\end{table}

\subsubsection{Dependence of the performance of the new method on the ion energy}$\newline$
$\ $In this section, we investigated the energy-dependence of performance on low-velocity carbon ions for the new method. Here, carbon energies of 30, 60, and 100 keV were evaluated.
The detected event densities were (1.03 $\pm$ 00.6)$\times$$10^7/cm^2$, and (0.68 $\pm$ 0.06)$\times$$10^7/cm^2$ for 60 and 30 keV, respectively, and 87$\%$ and 80$\%$ remained after the brightness cut in each energy, respectively. The decrease in detected event density in the case of 30 keV may be due to the existence of escape events, which were recoiled at the NIT surface region.

\begin{figure}[htbp]
\begin{center}
\includegraphics[width=90mm]{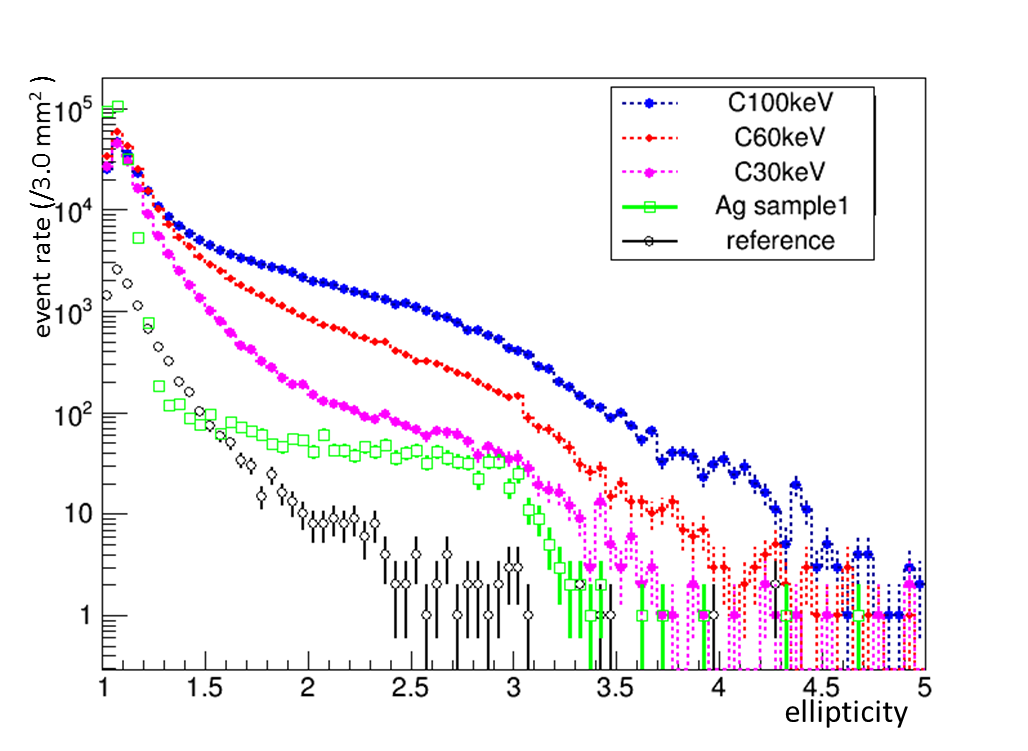}
\caption{Ellipticity distribution of the carbon ion samples, the reference sample and Ag sample 1 reconstructed by the new method.} 
\label{fig:Celli}
\end{center}
\end{figure}

The ellipticity distribution of each sample is shown in Fig.~\ref{fig:Celli}. It can be seen that the higher-ellipticity events increases as the energy increases.
The CC events in Ag sample 1, for which the event number density is comparable to that of the ion samples as shown before, has a different distribution shape, and it can be seen that the signal events of ion tracks are distributed with the ellipticity greater than 1.3.

\begin{figure}[htbp]
\begin{center}
\includegraphics[width=160mm]{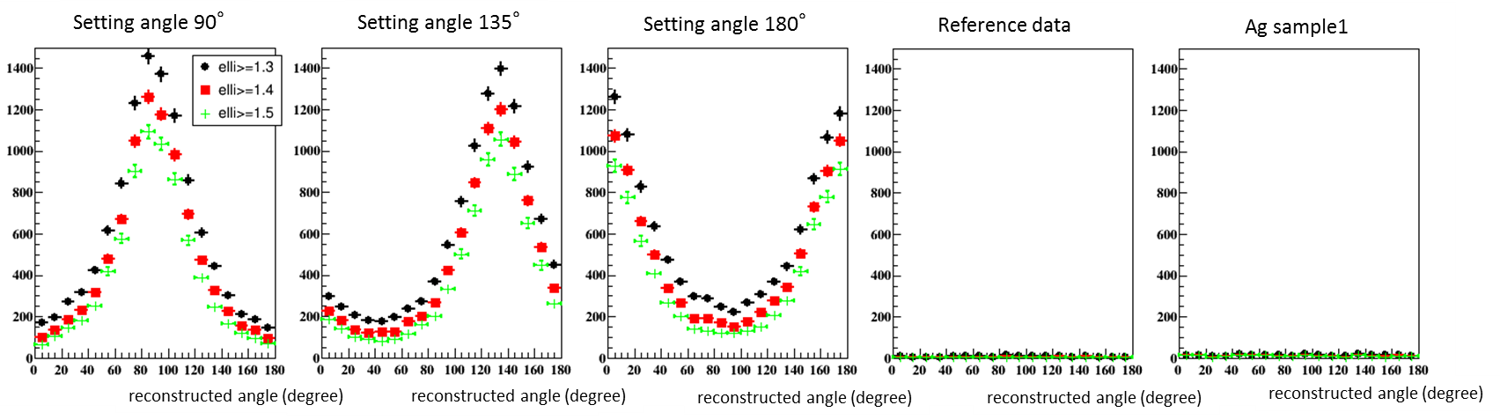}
\caption{Reconstructed angle distribution of the carbon 100 keV ions with ellipticity thresholds, 1.3, 1.4, and 1.5. The right two graphs show the results of the reference and Ag sample 1 with the same ellipticity threshold and scanning area. } 
\label{fig:100_angle}
\end{center}
\end{figure}

\begin{figure}[htbp]
\begin{center}
\includegraphics[width=160mm]{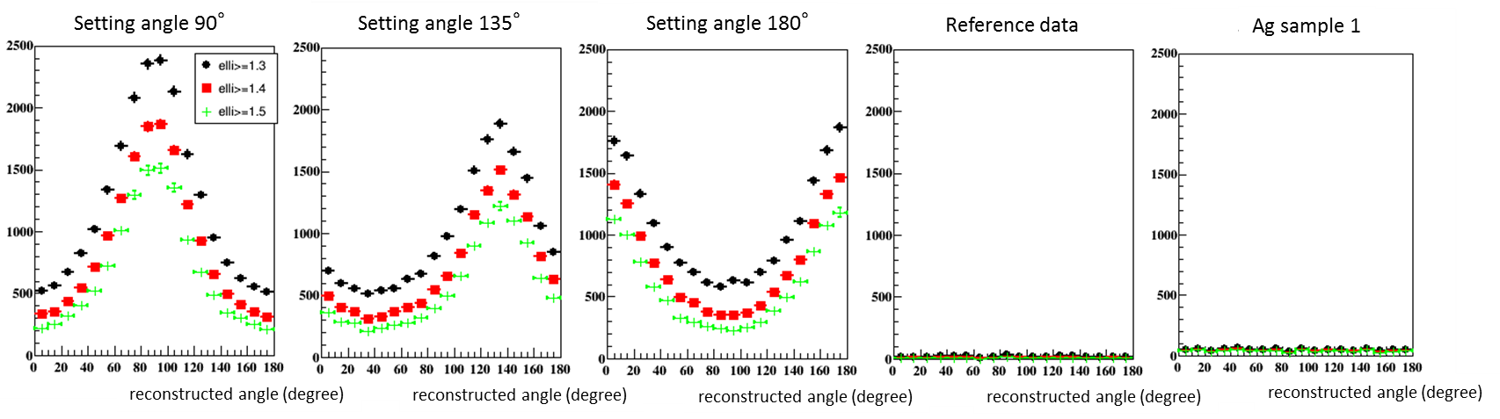}
\caption{Reconstructed angle distribution of the carbon 60 keV ions with different ellipticity thresholds, 1.3, 1.4 and 1.5.} 
\label{fig:60_angle}
\end{center}
\end{figure}

\begin{figure}[htbp]
\begin{center}
\includegraphics[width=160mm]{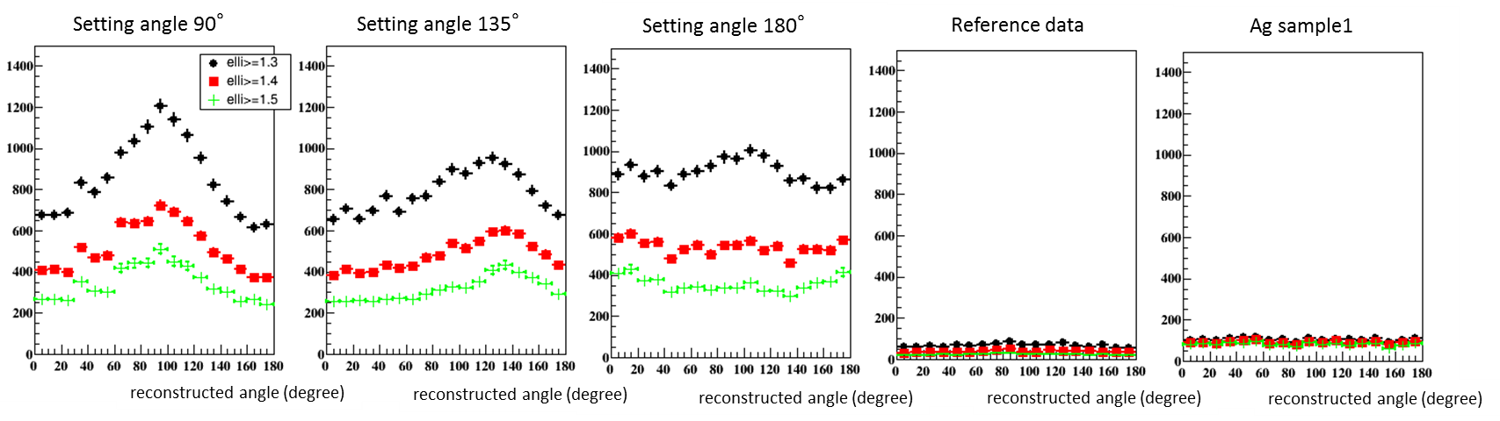}
\caption{Reconstructed angle distribution of the carbon 30 keV ions with different ellipticity thresholds, 1.3, 1.4 and 1.5.} 
\label{fig:30_angle}
\end{center}
\end{figure}

\begin{figure}[htbp]
\begin{center}
\includegraphics[width=160mm]{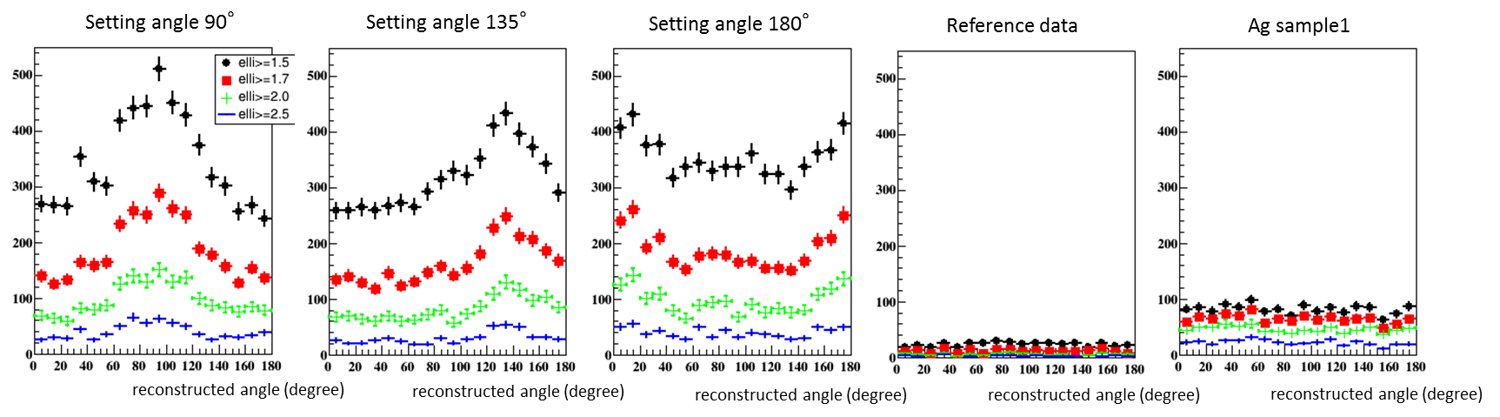}
\caption{Reconstructed angle distribution of the carbon 30 keV ions with different ellipticity thresholds, 1.5, 1.7, 2.0, and 2.5.} 
\label{fig:30_angle2}
\end{center}
\end{figure}

The reconstructed angular distributions by the new method for the cases of carbon 100, 60, and 30 keV are shown in Fig.~\ref{fig:100_angle} to Fig.~\ref{fig:30_angle2}. To maintain the number of events that can be used for evaluation of the angular resolution, i.e., the events with larger ellipticity, the scanning area for each energy was set at 1.5 mm$\times$1.5 mm, 1.2 mm$\times$1.2 mm, 0.6 mm$\times$0.6 mm for the case of 30, 60, and 100 keV, respectively.

In Fig.~\ref{fig:100_angle} and Fig.~\ref{fig:60_angle} (cases of 100 and 60 keV), clear peaks at the right-angle position are recognized with ellipticity cuts of elli $\geq$1.3, elli $\geq$ 1.4 and elli $\geq$1.5. No apparent difference is observed between the different setting angles. In contrast, in the 30 keV data in Fig.~\ref{fig:30_angle}, the shape of the distribution is changed depending on the installation direction. In the 180$^\circ$ setting in particular, it is not possible to detect the incident direction, especially in the case of a low ellipticity threshold. For the reference, in Fig.~\ref{fig:30_angle2}, the distributions with higher threshold (elli $\geq$1.5, elli $\geq$ 1.7, elli $\geq$ 2.0 and elli $\geq$ 2.5) are shown. The peak becomes recognizable even at 180$^\circ$ setting when the threshold becomes higher. 

\begin{figure}[htbp]
\begin{center}
\includegraphics[width=160mm]{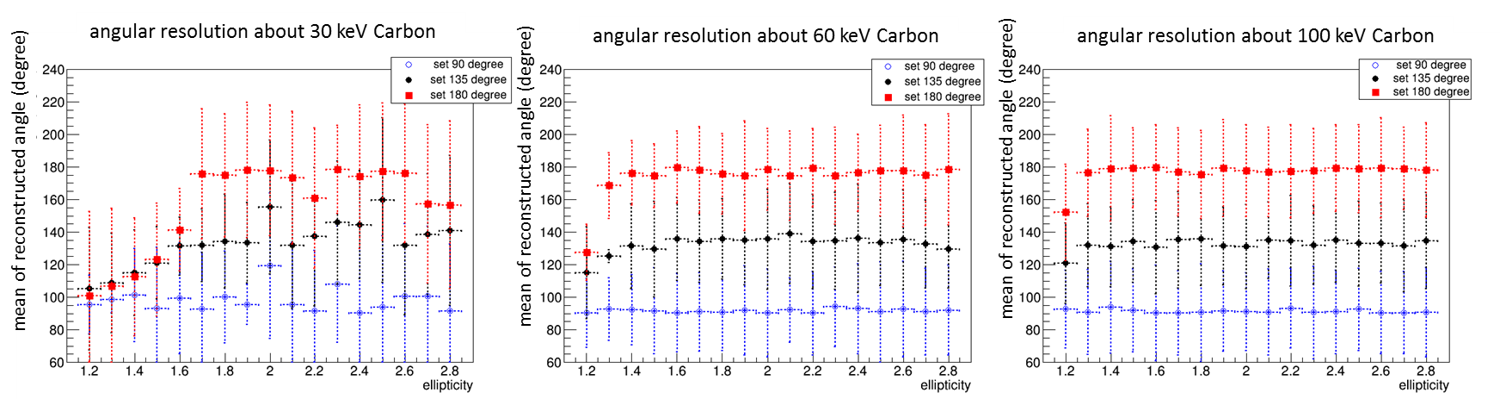}
\caption{Track angle reconstruction quality about low-velocity carbon by elliptical analysis. The average of the reconstructed angles for each binned sample is plotted as a dot, and the RMS is shown as the bar extending from the dot.} 
\label{fig:Cangle}
\end{center}
\end{figure}

To study the origin of the effect, the events were sub-divided into ellipticity intervals as in the case of Fig.~\ref{fig:C100_RMS}, and  angle reconstruction was performed for each binned sample. The results are shown in Fig.~\ref{fig:Cangle}. The expression is slightly different from Fig.~\ref{fig:C100_RMS}. The average of the reconstructed angles for each binned sample is plotted as a dot, and the RMS is expressed as the length of the bar extending from the dot.
The reconstruction seems good for ellipticity greater than 1.7, 1.4 and 1.3 in the case of 30, 60, and 100 keV, respectively. However, at low ellipticity, especially below 1.5 in the 30keV case, the average of the recognized angle tends to come to approximately 90$^\circ$ regardless of the setting angle. This effect may be caused by several factors, for example, one candidate is the micro-vibration of the scanning system that accompanies its movement, and the other is the aberration of the optics. The intrinsically wider angle distribution caused by the larger scattering with short track length about carbon at 30 keV may be easily influenced by these effects. To suppress the effect of the vibration, it is necessary to shorten the image capturing time. This is possible by using a pulsed light source and/or camera with a higher frame rate. These are subjects for future improvement. 

Using the current scanning system, we must apply the ellipticity threshold depending on the target energy. As seen in Fig.~\ref{fig:Cangle}, ellipticity thresholds of 1.7, 1.4, and 1.3 are selected for 30, 60, and 100 keV, respectively. As a result, the detection efficiency for carbon ions becomes 1.7$ \pm$ 0.1 $\%$, 13.1$\pm$ 0.1$\%$ and 29.7 $\pm$ 0.7 $\%$ and the angle reconstruction accuracy is 41$^\circ$, 28$^\circ$, and 27$^\circ$ for the respective energies of 30, 60, and 100 keV, as shown in Table~\ref{tb:Csummary}. 

It should be emphasized that this is the first report of successful angle reconstruction of carbon 30 keV ions in the world, although the detection efficiency and angular accuracy for 30 keV are 1.7 $\%$ and 45$^\circ$, respectively.

\begin{table}[htbp]
\centering
\begin{tabular}{c|c|c|c|c}
\hline
sample & ellipticity threshold & setting angle & $R$ & RMS \\ 
\hline
& & 90$^\circ$&1.8 $\pm$ 0.1 ($\%$) & 38 $\pm$ 8$^\circ$\\ 
carbon 30 keV & 1.7 & 135$^\circ$&1.6 $\pm$ 0.1 ($\%$) & 40 $\pm$ 9$^\circ$\\
& & 180$^\circ$&1.8 $\pm$ 0.1 ($\%$) & 44 $\pm$ 6$^\circ$\\
\hline
& & 90$^\circ$&13.2 $\pm$ 0.1 ($\%$) & 27 $\pm$ 4$^\circ$\\
carbon 60 keV & 1.4 & 135$^\circ$&13.1 $\pm$ 0.1 ($\%$) & 29 $\pm$ 7$^\circ$\\
& & 180$^\circ$&13.0 $\pm$ 0.1 ($\%$) & 28 $\pm$ 5$^\circ$\\
\hline
& & 90$^\circ$&30.5 $\pm$ 0.5 ($\%$) & 27 $\pm$ 2$^\circ$\\
carbon 100 keV & 1.3 & 135$^\circ$&29.7 $\pm$ 0.5 ($\%$) & 26 $\pm$ 2$^\circ$\\
& & 180$^\circ$& 28.8 $\pm$ 0.4 ($\%$) & 28 $\pm$ 2$^\circ$\\
\hline
\end{tabular}
\caption{Summary of the detection ability about carbon 30, 60, and 100 keV by the new elliptical analysis method. The recognition efficiency ($R$) and the angular resolution (RMS) for each ellipticity threshold are summarized.}
\label{tb:Csummary}
\end{table}
\clearpage

\section{Conclusion and discussion}
~~To detect and reconstruct sub-micron range recoil tracks recorded in the NIT with PTS2 read-out system, we have developed a new elliptical analysis method using DFT and brightness moment. 
The new method decreased the pixel digitization effects, realized a uniform response in the angle space, and improved angle reconstruction accuracy. Though these improvements, we succeeded in lowering the ellipticity threshold to 1.7, 1.4, and 1.3 for 30, 60, and 100 keV, respectively. This resulted in, improved detection efficiency for carbon ions of 1.7 $\pm$ 0.1$\%$, 13.1 $\pm$ 0.1$\%$ and 29.7 $\pm$ 0.7$\%$, and improved angle reconstruction accuracy of 41$^\circ$, 28$^\circ$, and 27$^\circ$ for these energies, respectively. The detection efficiency at 100 keV is approximately1.5 times higher than the old method, and the success of the beam direction reconstruction of the 30 keV carbon beam is the first of its type in the world. 

In addition, by investigating the poor angular resolution at 30 keV, we found subjects of our microscope system for further improvements (i.e., aberration of the lens and vibration of the scanning stage). 
Although this method is under-developed, we demonstrated the capability of WIMPs detection with directional sensitivity and the potential to verify the DAMA~\cite{DAMA} experimental results, and to search around WIMPs of 10 GeV/$c^2$ mass, which are not reachable by conventional directional dark matter experiments \cite{Directional}. 
Subsequently, the low detection efficiency in elliptical analysis, can be improved to change the distribution and the size of track-grains, which affected the optical shape. We are also studying a track detection simulator that combines a track generator including NIT crystal arrangements with optical simulation.
 
\vskip2pc

\section*{Acknowledgment}    
This work was supported by JSPS KAKENHI Grant Numbers JP18H03699, JP19H05806 and by "Nanotechnology Platform"(F-19-NU-006) of the Ministry of Education, Culture, Sports, Science and Technology (MEXT), Japan and Nagoya University Nano fabrication Platform.

\end{document}